# What does it all mean? Capturing Semantics of Surgical Data and Algorithms with Ontologies

Darko Katić[1], Maria Maleshkova[2], Sandy Engelhardt[3], Ivo Wolf[3], Keno März[3], Lena Maier Hein[3], Marco Nolden[3], Martin Wagner[4], Hannes Kenngott[4], Beat Peter Müller-Stich[4], Rüdiger Dillmann[1], and Stefanie Speidel[1]

Institute for Anthropomatics and Robotics, Karlsruhe Institute of Technology[1]
Institute of Applied Informatics and Formal Description Methods, Karlsruhe Institute of Technology[2]
Division of Medical and Biological Informatics, German Cancer Research Center (DKFZ), Heidelberg[3]
Department of General, Abdominal and Transplantation Surgery, University of Heidelberg[4]
`darko.katic@kit.edu`

**Abstract.** Every year approximately 234 million major surgeries are performed, leading to plentiful, highly diverse data. This is accompanied by a matching number of novel algorithms for the surgical domain.
To garner all benefits of surgical data science it is necessary to have an unambiguous, shared understanding of algorithms and data. This includes inputs and outputs of algorithms and thus their function, but also the semantic content, i.e. meaning of data such as patient parameters. We therefore propose the establishment of a new ontology for data and algorithms in surgical data science. Such an ontology can be used to provide common data sets for the community, encouraging sharing of knowledge and comparison of algorithms on common data. We hold that this is a necessary foundation towards new methods for applications such as semantic-based content retrieval and similarity measures and that it is overall vital for the future of surgical data science.

## 1 Problem Statement

An estimated 234 million major surgeries are performed every year world-wide [1]. The data created during these surgeries is plentiful and highly diverse. Intraoperative information artifacts alone range from instrument applications and trajectories over images from microscopes, endoscopes or ultrasound. Postoperatively further diagnoses, plannings and outcomes need to be included. The technical challenge is the heterogeneity of the data. Time variant states (e.g. time-series coming from laboratory results or follow-up images of disease progression), processes (e.g. intraoperative workflows) and device states need to be considered in a combined analysis.

The amount of data is matched by the availability of algorithms. There is a vast amount of approaches for surgical data science applications ranging from preoperative decision support over intraoperative assistance generation and workflow optimization to post-operative skill assessment. Current algorithms use a variety of possible inputs. Even algorithms solving similar problems often use different data. Amongst others, this is evident in the modeling of intraoperative workflows where there is no ultimate agreement on relevant features [2].

A deep, clear and machine-readable understanding of the underlying data is also important for machine learning. This is best articulated by the No-Free-Lunch-Theorem [3]. The theorem states that, under very general conditions, there is no universally better algorithm for machine learning. Instead, the algorithm to be used needs to be chosen specifically for the data and the problem at hand. A clear, formal description of data and algorithms is therefore necessary.

The quantity, heterogeneity and interconnectedness of algorithms and data are salient challenges to surgical data science. To truly realize its potential of holistic analysis, a language for data and algorithms needs to be found. A shared understanding on the semantic level is necessary to cope with the large amount of data and algorithms. Especially in a collaborative community this is important as it supports sharing of data and helps avoid misunderstandings.

To this end, we propose the development of an ontology for algorithms and data in surgery. Specifically, we aim to capture semantics, e.g. what the data is about. The usefulness of such ontologies is apparent in other domains, e.g. data mining [4]. Current medical ontologies like OntoSPM [5], the Foundational Model of Anatomy (FMA) [6] and SNOMED-CT [7] focus only on medical knowledge itself, yet not on data and algorithms, while OntoDM [4] is not directly applicable to medical use-cases without further development.

## 2 Potential Benefits and Uses

Assuming sufficient acceptance, such an ontology offers a community-wide consensus on terms, algorithms and data structures. This greatly facilitates comparison of proposed approaches and encourages sharing of knowledge and data. These steps are not only desirable but salient staples of good scientific practice.

The ontology inherently models provenance, i.e. origin of data. This allows to estimate trust to arbitrate between conflicting assessments. For instance it can be queried whether a segmentation was obtained purely algorithmically, by an experienced clinician or a novice one.

On a methodological level, such an ontology supports the development of novel data interpretation techniques. It can be used for content-based retrieval of images or cohorts. Furthermore, it lends itself to use in similarity based retrieval over different modalities. For machine learning applications, such an ontology is very beneficial as it allows fine-granular, semantic selection of training data based on the semantic description. It also aids in the selection of machine learning approaches as it allows a better understanding of processes and data generation.

Further applications include easier search for algorithms tailored to specific purposes, improved understanding as well as documentation. Reasoning techniques can be used to provide even more advanced functions such as automatic proposals of algorithms or even of entire pipelines in specific use-cases [8].

## 3   Requirements and Challenges

The development of such an ontology is challenging. For once, it must be accepted by the research community. The problem is including many stakeholders during the development and finding a consensus. Furthermore the proposed ontology needs to include relevant ontologies such as SNOMED-CT [7] and FMA [6]. This is crucial for interoperability with system using such established ontologies. Content-wise, the ontology needs to contain rich semantics and allow for reasoning. A mere taxonomy is insufficient. The associated challenge is to gather, combine and formalize expert knowledge from clinicians and technicians.

To reap the benefits of the ontology, real-world data sets and algorithms must be described with it. This requires considerable effort. Methodically, the annotation can be created by a mix of manual work, crowd-sourcing and automatic labeling. Thus it can be parallelized and partly automated.

Due to these challenges and the large scope, we propose to first create a mid-level ontology to ensure general agreement. It can then be refined to application ontologies for specific purposes. Because of the large amount of effort and coordination required, widespread collaboration is mandatory.

## 4   Technical Feasibility Analysis

Such an ontology is difficult yet feasible to attain. Reusability and adherence to standards can be ensured using an upper-ontology. Upper-ontolgies define general terms which are further specialized for concrete domains. The main advantage is that different ontologies built with the same upper-ontology can be connected rather easily, as they already share a foundation. Furthermore they improve the quality of ontologies offering a well thought-out basic categorization.

We propose the use of Basic Formal Ontology (BFO) as the upper-ontology [9]. BFO is mature and used in many existing (bio-)medical ontologies. Furthermore, we aim to include the Information Artefact Ontology [10] and the the Ontology of Biomedical Investigation [11].

An idea of the ontology is shown in Fig. 1. The example from the domain of heart surgery illustrates the representation of a DICOM-file showing the mitral valve in the long axis plane, captured using ultrasound with the surface of the valve segmented via an implementation of a segmentation algorithm. Algorithms and their implementations are modeled separately. The processing of data is modeled by expressing the process of execution of the implementation with inputs and outputs modeled as participants. For applications needing user input, e.g. interactive segmentations, additional participant can be added. This way the creation of data and thus provenance is modeled.

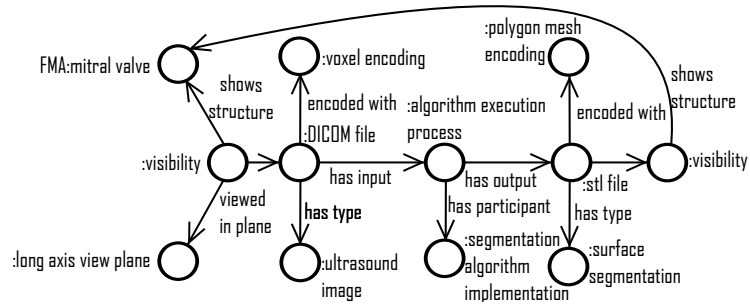

**Fig. 1.** Representation of a file showing the mitral valve in the long axis plane, captured using ultrasound used to create a segmentation of the mitral valve surface.

Overall, we believe the development of the ontology is a feasible and very fruitful endeavour.

*Acknowledgements* The present research was supported by the "SFB TRR 125" funded by the DFG.